\documentclass{article}
\usepackage{amsmath,graphicx,mlspconf, url} 

\usepackage[capitalise]{cleveref}

\usepackage[labelsep=period,labelfont=bf]{caption}
\usepackage{subcaption}
\usepackage{amsmath}
\usepackage{wrapfig}

\usepackage{xcolor}

\expandafter\def\expandafter\normalsize\expandafter{%
    \normalsize%
\setlength\abovedisplayskip{2pt}%
\setlength\belowdisplayskip{8pt}%
\setlength\abovedisplayshortskip{-8pt}%
    \setlength\belowdisplayshortskip{2pt}%
}

\toappear{2024 IEEE International Workshop on Machine Learning for Signal Processing, Sept.\ 22--25, 2024, London, UK}

\title{TINA: Acceleration of Non-NN Signal Processing Algorithms Using NN Accelerators}

\name{%
    Christiaan Boerkamp$^{\star}$%
    \qquad Steven van der Vlugt$^{\dagger}$%
    \qquad Zaid Al-Ars$^{\star}$
}
\address{%
    $^{\star}$Delft University of Technology, Mekelweg 4, 2628 CD, Delft, Netherlands \\%
    $^{\dagger}$ASTRON, Oude Hoogeveensedijk 4, 7991 PD Dwingeloo, Netherlands%
}

\begin{document}

\maketitle

\begin{abstract}
This paper introduces TINA, a novel framework for implementing non Neural Network (NN) signal processing algorithms on NN accelerators such as GPUs, TPUs or FPGAs. The key to this approach is the concept of mapping mathematical and logic functions as a series of convolutional and fully connected layers. By mapping functions into such a small sub stack of NN layers, it becomes possible to execute non-NN algorithms on NN hardware (HW) accelerators efficiently, as well as to ensure the portability of TINA implementations to any platform that supports such NN accelerators.   
Results show that TINA is highly competitive compared to alternative frameworks, specifically for complex functions with iterations. For a Polyphase Filter Bank use case TINA shows GPU speedups of up to 80x vs a CPU baseline with NumPy compared to 8x speedup achieved by alternative frameworks. The framework is open source and publicly available at {\small\url{https://github.com/ChristiaanBoe/TINA}}.

\end{abstract}
\section{Introduction}
\label{sec:intro}

In recent years, there has been a rapid increase in the number of specialized NN accelerators, either as standalone HW (e.g., TPUs) or as accelerated components in existing HW (e.g., GPUs or FPGAs). This creates a lucrative opportunity for using such accelerators to boost the performance of non-NN algorithms, in the same way that GPUs are used to accelerate non-graphics algorithms. 

Existing research performed on this topic is limited. 
One example is JAX, a machine learning framework for transforming mathematical functions and mapping them on NN HW~\cite{jax2018github}. However, JAX is mainly focused on making ML programming more intuitive, rather than providing a general API to run non-NN algorithms on NN HW. Other efforts are either limited to a specific application (e.g., brain simulation~\cite{christos_tpu}), or limited to a specific HW platform (e.g., GPUs~\cite{CuPy} or FPGAs~\cite{joost_frame}).

This paper introduces a novel framework, referred to as TINA\footnote{TINA stands for "This Is Not AI"}, to reduce the complexity of implementing non-NN applications on NN accelerators, thereby ushering in the era of GPNPUs (general-purpose neural processing units). In this paper, we introduce the idea of representing signal processing functions as a series of convolutions and fully connected layers. TINA makes use of the strengths of NN HW designed to accelerate NN layers, such as matrix operations. By reinterpreting traditional mathematical and logic functions as NN layers, we facilitate a form of computation that is intrinsically compatible with NN accelerators. As its input programming language, TINA uses Python, one of the most widely used programming languages in data science~\cite{joost_bigdata}.

\begin{table}
    \centering
    \caption{Signal processing functions implemented in this paper and the TINA building blocks used to implement them}
    \label{tab:implemented_operators}
    \begin{tabular}{@{}lll@{}}
    \hline
         Function & Building blocks & Section \\
         \hline \hline
         Elementwise matrix mult.\ & Depthwise conv.\ & \cref{sec:elementmult} \\ \hline
         Matrix-matrix mult.\ & Depthwise conv.\ & \cref{sec:matmatmult} \\ \hline
         Elementwise matrix add & Depthwise conv.\ & \cref{sec:elementadd} \\ \hline
         Summation& Fully conn.\ layer & \cref{sec:Summation} \\ \hline
         DFT & Pointwise conv.\ & \cref{sec:DFT} \\ \hline
         Inverse DFT & Pointwise conv.\  & \cref{sec:IDFT} \\ \hline
         FIR filter&  Standard conv.\ & \cref{sec:FIRfilt} \\ \hline
         Unfolding algorithm& Standard conv.\ & \cref{sec:slidingwidnowalgorithmn} \\ \hline
    \end{tabular}
\end{table}

TINA provides a number of unique advantages, such as 
 
ensuring the portability of TINA implementations to any HW platform that supports NN accelerators.  
Furthermore, TINA allows non-NN algorithms to be optimized using methods previously exclusive to NNs (e.g., pruning libraries or automatically optimized quantization using quantize aware training libraries such as Brevitas \cite{Xilinx}). 

TINA consists of two main components: 1.\ basic building blocks, and 2.\ APIs that use the building blocks to implement various functions. \cref{tab:implemented_operators} provides an overview of the functions discussed in this paper and the building blocks used to implement them.

\section{Building blocks of TINA}
\label{sec:Buildingblocks}
The building blocks of TINA can be divided into four NN layers: standard convolution, depthwise convolution, pointwise convolution and fully connected layers. These building blocks will be used later in the paper to implement signal processing functions. We use these specific four NN layers since they are the most widely used layers in ML applications \cite{Albawi2017Understanding, sym14040658}, thereby making them well-supported by accelerators, and easily portable to alternative hardware platforms.

\subsection{Standard convolution}
\label{sec:stanconv}

Convolutions are critical in deep learning for feature extraction from various types of input data, such as images and audio. This process involves sliding a kernel across the input data, conducting elementwise multiplication with the input values covered by the kernel, and summing these products to generate output feature maps, as depicted in \cref{math:mathconv}.
\begin{equation}
\label{math:mathconv}
\begin{split}
&O(h,w, c_{out})=  b(c_{out}) + \\ &\sum^{C_{in}}_{c_{in}}\sum^{M}_{m}\sum^{N}_{n}I(h+m,w+n,c_{in}) \cdot K(m,n,c_{in}, c_{out})\
\end{split}
\end{equation}

where $O$ is the output, $I$ is the input, $K$ is the kernel, $c_{in}$ and $c_{out}$ are indices representing the input and output channels, respectively, while $b$ is the bias. TINA leverages PyTorch to enable precise customization of convolutional layers. This customization includes controlling the dimensions of both the input matrix, represented as ($T$, $C_{in}$, $H$, $W$), and the output matrix, represented as ($T$, $C_{out}$, $H$, $W$), where $T$ is the batch size, $C$ denotes the number of channels, and $H$ and $W$ are the height and width of the input/output, respectively. 

PyTorch provides a variety of parameters to allow for extra controls over the convolution. For example, we can use \textbf{groups} to define blocks of connections between input and output channels. Other notable parameters include \textbf{stride}, determining the kernel movement steps, and \textbf{padding}, adjusting the input border size.

\subsection{Depthwise convolution}
\label{sec:Depthwiseconv}
Depthwise convolution represents a mechanism within convolutional neural networks for dissecting spatial relationships in input data. Distinct from standard convolutional layers, which convolve across both input channels and spatial dimensions, depthwise convolution applies the corresponding channel of the kernel to each input channel independently, resulting in an output with the same number of channels as the input. This operation is represented in \cref{math:depthconv}.

\begin{equation} \label{math:depthconv}
\begin{split}
&O(h,w,c_{out})= b(c_{out}) + \\
&\sum^{M}_{m}\sum^{N}_{n}I(h+m,w+n,c_{out}) \cdot K(m,n,c_{out})
\end{split}
\end{equation}

Depthwise convolution causes the kernel to slide across the  input data, executing elementwise multiplications between the kernel and input values, with the sums producing the output. This procedure is independently executed for each channel, thereby generating distinct feature maps for every channel.

\subsection{Pointwise convolution}
\label{sec:pointconv}

Pointwise convolutions, also known as $1 \times 1$ convolutions, are a special case of standard convolutions where a kernel of dimensions $1 \times 1$ is used to process individual elements across the channels of input data, which allows for the mixing of channel information. This operation is represented in \cref{math:pointwise_equation}. 
{\small
\begin{equation} \label{math:pointwise_equation}
O(h,w, c_{out}) = b(c_{out}) + \sum^{C_{in}}_{c_{in}}{\ } I(h,w,c_{in}) \cdot K(c_{in}, c_{out})
\end{equation}
}

\subsection{Fully connected layer}
\label{sec:fullconlay}
A fully connected layer, also known as a dense layer or linear layer, is a commonly-used layer in many neural network architectures. Its core role is to transform input data by applying a linear operation, typically followed by a non-linear activation function to introduce non-linearity into the model. The operation performed by a fully connected layer involves computing the output from the input through matrix multiplication with the kernel matrix, and adding a bias term $b$, as shown in \cref{math:fullyconnected}.
\begin{equation} \label{math:fullyconnected}
O(c_{out})= b(c_{out}) + \sum^{C_{in}}_{c_{in}} I(c_{in}) \cdot K(c_{in},c_{out})
\end{equation}

\section{Arithmetic functions mapping}
\label{sec:OPmapMath}
In this section, we show how we can compute various arithmetic functions using the TINA building blocks discussed in Section~\ref{sec:Buildingblocks}.

\subsection{Elementwise multiplication}
\label{sec:elementmult}

An elementwise matrix multiplication can be implemented using a depthwise convolution. As shown in \cref{math:DepthwiseConvOrig}, a depthwise convolution applies the corresponding channel of the kernel to each input channel independently, causing the kernel to slide across the input matrix, executing elementwise multiplications between the kernel and input values, with the sums producing the output. 

\begin{equation}
\begin{split} \label{math:DepthwiseConvOrig}
& O(h,w,c_{out}) = b(c_{out}) + \\ 
&\sum^{M}_{m}\sum^{N}_{n}I(h+m,w+n,c_{out}) \cdot K(m,n,c_{out})
\end{split}
\end{equation}

In order to have this equation represent an elementwise matrix multiplication, we first set the bias $b(c_{out})=0$. In addition, we convert the width and height of both the input matrix and the kernel into 1$\times$1 matrices and reshape the elements of these two matrices into vectors along the channel axis where $C_{out}=H \times W$, resulting in \cref{math:ElementwisematReduced}. 
\begin{align}
O(0,0,c_{out}) 
&= \sum^{0}_{0}\sum^{0}_{0}I(0+m,0+n, c_{out}) \cdot K(m,n,c_{out}) \nonumber \\
&= 
I(0,0,c_{out}) \cdot K(0,0,c_{out}) \nonumber \\
\Rightarrow O(c_{out}) & =I(c_{out}) \cdot K(c_{out}) \label{math:ElementwisematReduced}
\end{align}

\subsection{Matrix-matrix multiplication}
\label{sec:matmatmult}

A matrix-matrix multiplication is a common operation in linear algebra and can be seen as a combination of a summation and a pointwise multiplication as shown in \cref{math:matmatmultexample}, where $X(m,l)$ represents the first input matrix X with dimensions $M$ and $L$, $Y(l,n)$ represents the second input matrix Y with dimensions $L$ and $N$, and $Z(m,n)$  represents the output matrix Z with dimensions $M$ and $N$.
\begin{equation}
\label{math:matmatmultexample}
Z(m,n) = \sum_{l}^{L} X(m,l) \cdot Y(l,n)
\end{equation}
We can represent the matrix-matrix multiplication using a pointwise convolution  
by taking the following additional steps. We start with setting the bias of the pointwise convolution $b(c_{out})$ to zero  
resulting in \cref{math:pointwise_equation_nobias}.
\begin{equation} \label{math:pointwise_equation_nobias}
O(h,w, c_{out})=  0 + \sum^{C_{in}}_{c_{in}} I(h,w,c_{in}) \cdot K(c_{in}, c_{out})
\end{equation}

In order for \cref{math:pointwise_equation_nobias}  to represent a matrix-matrix multiplication, we apply the following transformation as shown in \cref{math:pointwise_equation_matmat_mult}, by reshaping our input and output matrices with dimensions $H$ and $W$ into vectors with dimension $1 \times M$ where $M = H \times W$, and making $C_{in} = L$ and $C_{out} = N$. This in turn is equal to our original definition of a matrix-matrix multiplication in \cref{math:matmatmultexample}.
\begin{align}
O(0, m, n) &=  \sum^{L}_{l} I(0, m, l) \cdot K(l, n) \nonumber \\ 
\Rightarrow O(m, n)&=\sum^{L}_{l} I(m, l) \cdot K(l, n) \label{math:pointwise_equation_matmat_mult}
\end{align}

\subsection{Elementwise addition}
\label{sec:elementadd}
Elementwise addition builds further upon the elementwise multiplication introduced in \cref{sec:elementmult}. By opting to set the weights of the kernel to an all-ones matrix and the bias $b(c_{out})$ of the convolution as one of the input matrices, we are able to transform \cref{math:ElementwisematReduced} to an elementwise addition as shown in \cref{math:Elementwiseadd}. Note that $I$, bias and $O$ in the equation represent vectors of size $C_{out} = H \times W$.
\begin{equation}
\label{math:Elementwiseadd}
\begin{split}
O(c_{out}) & = b(c_{out}) + I(c_{out}) \cdot K(c_{out}) \\ 
& = b(c_{out}) + I(c_{out}) \cdot \mathrm{1}(c_{out}) \\ 
\Rightarrow O(c_{out})  & = b(c_{out}) + I(c_{out})
\end{split}
\end{equation}

\subsection{Summation}
\label{sec:Summation}
The implementation of summation within TINA using the building blocks described in \cref{sec:Buildingblocks} can be  achieved through the utilization of a fully connected layer as described in \cref{sec:fullconlay}. By opting to set the number of output channels to 1, the weights of the kernel to an all-ones vector and the bias to a zero vector,  \cref{math:fullyconnected} can be transformed to a summation as shows in \cref{math:summation}. 
\begin{equation}
\label{math:summation}
\begin{split}
O(0) & = 0 + \sum^{C_{in}}_{c_{in}} I(c_{in}) \cdot 1(c_{in},0) \\ 
 \Rightarrow O & =  \sum^{C_{in}}_{c_{in}} I(c_{in})
\end{split}
\end{equation}

\section{Signal processing functions mapping}
\label{sec:OPmapsig}

\begin{figure*}
     \centering
     \begin{subfigure}[b]{0.24\textwidth}
         \centering
         \includegraphics[width=1\textwidth]{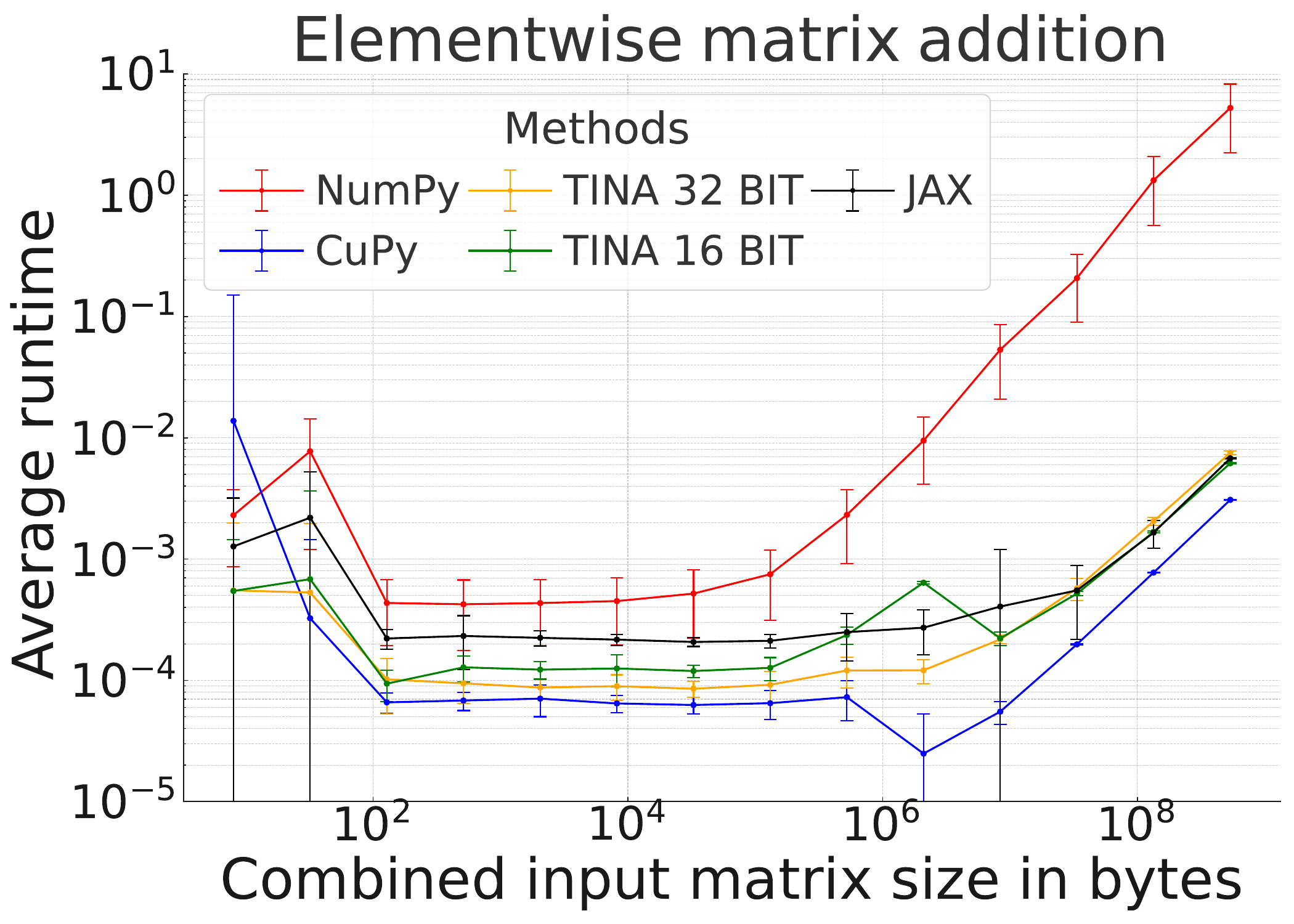}
         \caption{Elementwise matrix mult.}
         \label{fig:elementmatresults}
     \end{subfigure}
     \hfill
     \begin{subfigure}[b]{0.24\textwidth}
         \centering
         \includegraphics[width=1\textwidth]{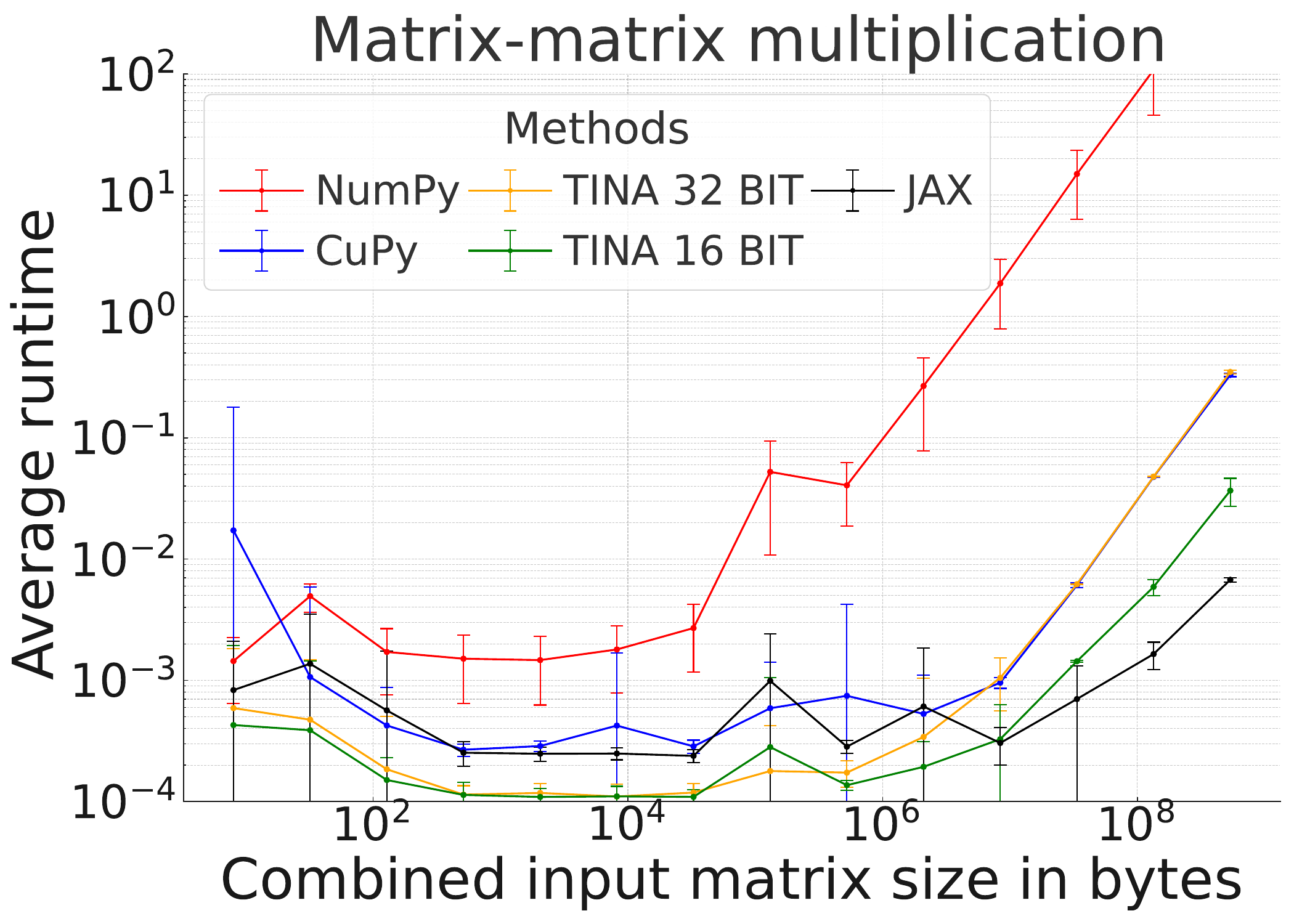}
         \caption{Matrix-matrix multiplication 
         }
         \label{fig:matmatresults}
     \end{subfigure}
     \hfill
     \begin{subfigure}[b]{0.24\textwidth}
         \centering
         \includegraphics[width=1\textwidth]{images/Elementwise_Matrix_Addition_Plot-2_1.pdf}
         \caption{Elementwise matrix addition}
         \label{fig:elementadditionresults}
     \end{subfigure}
     \begin{subfigure}[b]{0.24\textwidth}
         \centering
         \includegraphics[width=1\textwidth]{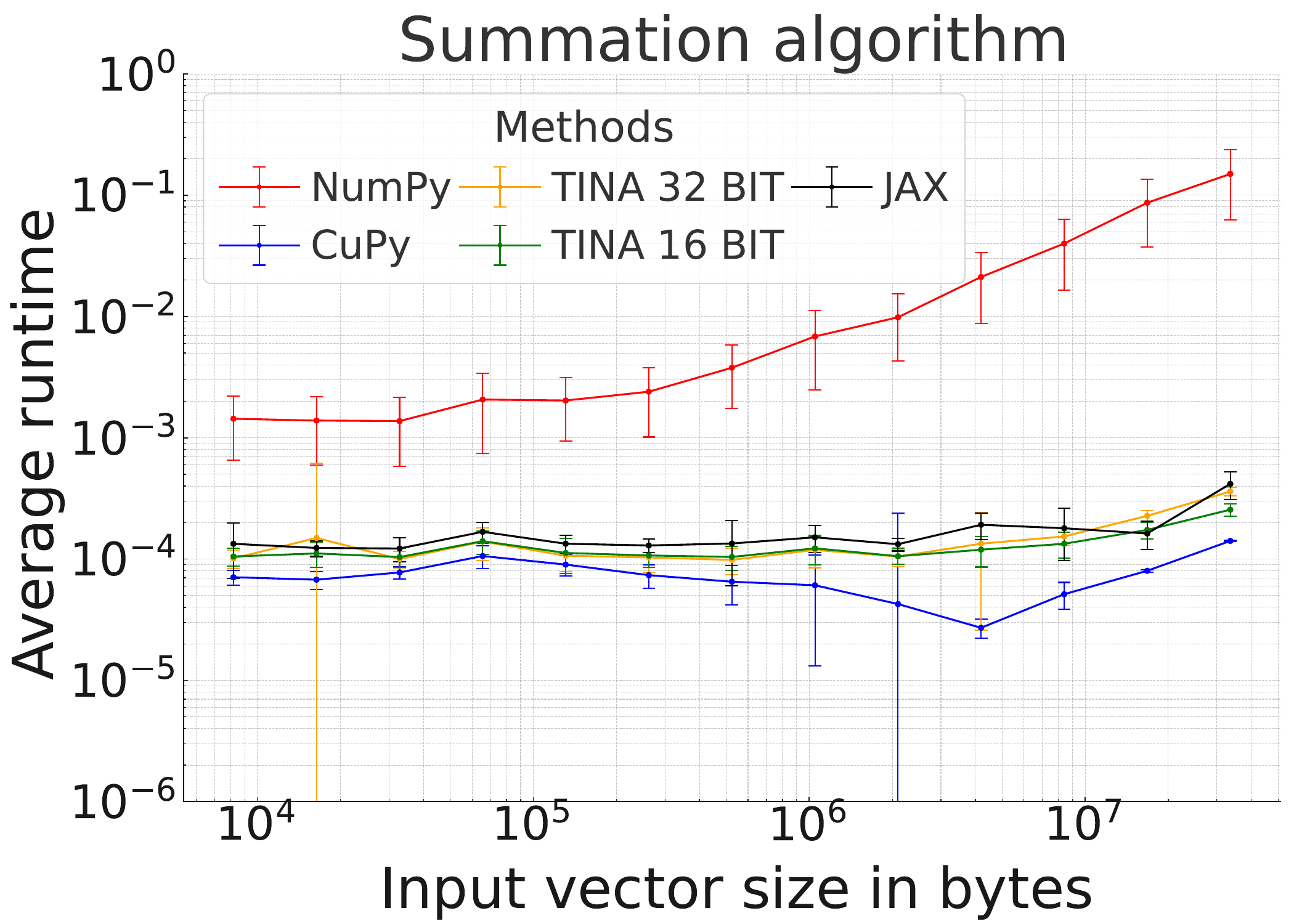}
         \caption{Summation algorithm}
         \label{fig:sumresults}
     \end{subfigure}
        \caption{Runtime of the arithmetic functions vs input size using TINA, NumPy, CuPy \& JAX 
        }
        \label{fig:arithresults}
\end{figure*}

\subsection{DFT}
\label{sec:DFT}
In this section, we show how to use TINA building blocks to implement a Discrete Fourier Transform (DFT). One way an input signal can be transformed towards the frequency domain is by matrix-matrix multiplying the input signal vector with a Discrete Fourier Matrix (DFM) \cite{Rao_Yip_2001} as seen in \cref{math:fourier}, with $X$ being the input signal and $F$ being the Fourier matrix.
\begin{equation}
\label{math:fourier}
Z(m,n) = \sum_{l}^{L} X(m,l) \cdot F(l,n)
\end{equation}
Using our matrix-matrix multiplication \cref{math:shortmatmatconv} as further detailed in \cref{sec:matmatmult} we can draw a direct parallel to \cref{math:fourier} by setting the kernel equal to the Fourier matrix.
\begin{equation}
\label{math:shortmatmatconv}
    O(m, n)=\sum^{L}_{l} I(m, l) \cdot K(l, n)
\end{equation}

\subsection{IDFT}
\label{sec:IDFT}
In this section, we show how to use TINA building blocks to implement an Inverse Discrete Fourier Transform (IDFT). One way an input Fourier signal can be transformed back towards the time domain is by matrix-matrix multiplying the input signal vector with an Inverse Discrete Fourier Matrix (IDFM) \cite{Rao_Yip_2001} as seen in \cref{math:invfourier}, with $Z$ as the input signal and $\mathit{IF}$ being the IDFM.
\begin{equation}
\label{math:invfourier}
X(i,j) = \sum_{k}^{K} Z(i,k) \cdot \mathit{IF}(k,j)
\end{equation}
Using our matrix-matrix multiplication \cref{math:shortmatmatconv} as further detailed in  \cref{sec:matmatmult} we can draw a direct parallel to \cref{math:invfourier} by setting the kernel equal to the IDFM.

\subsection{FIR filter}
\label{sec:FIRfilt}
In this section, we show how to implement a Finite impulse response (FIR) filter using a TINA layer.  Both an FIR filter and convolution involve a linear operation applied to a sequence of input values or signals as seen in \cref{math:FirFilter} representing an FIR filter. 
$a(k)$ represents the filter coefficients defining the weights (FIR taps) applied to each delayed input sample. These coefficients indicate how the input signal is weighted at different time indices. 
\begin{equation}
\label{math:FirFilter}
y(i)= 
\sum_{k}^{K} a(k) \cdot x(i-k)
\end{equation}

We can implement this equation using the standard convolution defined in \cref{sec:stanconv} by setting the dimensions $C_{in}$, $C_{out}$, $H$ and $M$ to 1. Then by setting the bias $b(c_{out})$ equal to zero, we get \cref{math:FIRasConv}.
\begin{align}
O(0,w, 0) &=  0 + \sum^{0}_{0} \sum^{0}_{0} \sum^{N}_{n} I(0+0,w+n,0) \cdot K(0,n,0,0) \nonumber \\ 
\Rightarrow O(w) &= \sum^{N}_{n} I(w+n) \cdot K(n) \label{math:FIRasConv}
\end{align}

Mathematically, an FIR filter represents a discrete-time system with a finite duration impulse response, typically characterized by a set of coefficients determining the filter's behavior. Using the standard convolution, we can simply set the weights of the kernel equal to these coefficients.

\subsection{Unfolding algorithm}
\label{sec:slidingwidnowalgorithmn}

An unfolding algorithm takes an input vector and produces an output matrix representing subsequences of the input with a successively increasing index. The unfolding algorithm can be represented using the \cref{math:slidingwindow}.
\begin{equation}
\begin{split}
\label{math:slidingwindow}
Y(i, j) &= X(i + j) \\
\end{split}
\end{equation}

For an input vector $X$ of length $I$ and for an unfolding window of width $J$, the output matrix $Y$ has dimensions of $(I-J+1) \times J$. As an example, for an input vector of length 4, $X = [1, 2, 3, 4]$, and a window of 2, the output $Y = [ [1, 2], [2, 3], [3,4] ]$, which is a matrix of dimensions $3 \times 2$.

We can implement this equation using the standard convolution defined in \cref{math:mathconv} (\cref{sec:stanconv}) by setting the $C_{in}$, $H$ and $M$ to 1. Then by setting the bias $b(c_{out})$ equal to zero, we get \cref{math:UnfoldasConv}.
{\small
\begin{align}
O(0,w,  c_{out}) &=  0 + \sum^{0}_{0} \sum^{0}_{0} \sum^{N}_{n} I(0+0,w+n,0) \cdot K(0,n,0, c_{out}) \nonumber \\ 
\Rightarrow  O(w,  c_{out}) &= \sum^{N}_{n}I(w+n) \cdot K(n,  c_{out}) \label{math:UnfoldasConv}
\end{align}
}

In order to reproduce \cref{math:slidingwindow}, 
first we make the kernel as defined in \cref{math:UnfoldasConv} a square matrix by setting the kernel dimensions $N$ = $C_{out}$. Then, we make the kernel weights equal to an identity matrix, which reproduces the input at the diagonal (i.e., when $n$ = $c_{out}$), else the input gets multiplied with 0. This results in \cref{math:unfoldconv}.

\begin{equation}
O(w,  c_{out}) = I(w+c_{out}) 
\label{math:unfoldconv}
\end{equation}

\section{Experimental results}
\label{sec:results}

\begin{figure*}
     \centering
     \begin{subfigure}[b]{0.24\textwidth}
         \centering
         \includegraphics[width=1\textwidth]{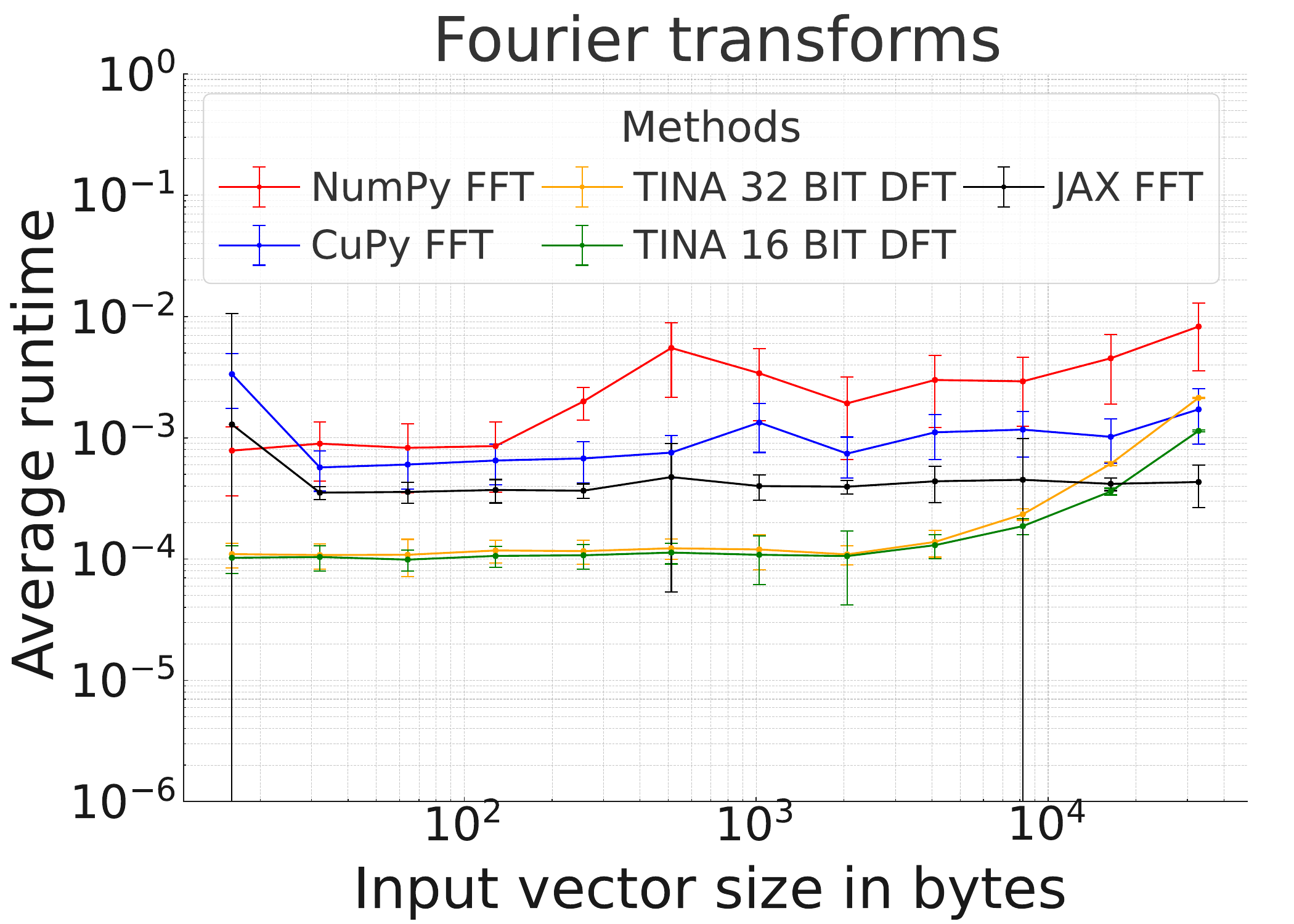}
         \caption{Fourier transform}
         \label{fig:DFTResults}
     \end{subfigure}
     \hfill
     \begin{subfigure}[b]{0.24\textwidth}
         \centering
         \includegraphics[width=1\textwidth]{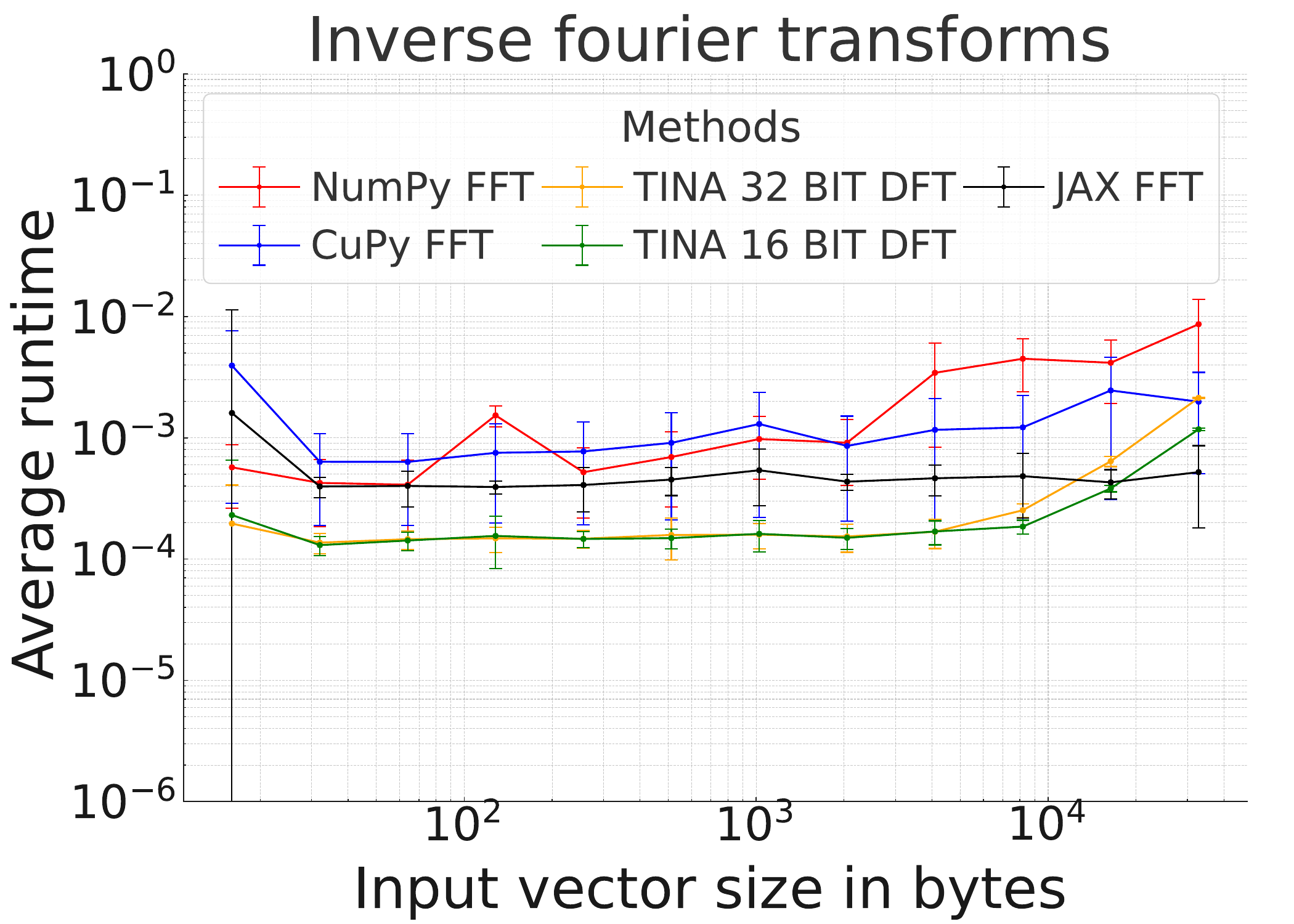}
         \caption{Inverse Fourier transform}
         \label{fig:IDFTResults}
     \end{subfigure}
     \hfill
     \begin{subfigure}[b]{0.24\textwidth}
         \centering
         \includegraphics[width=1\textwidth]{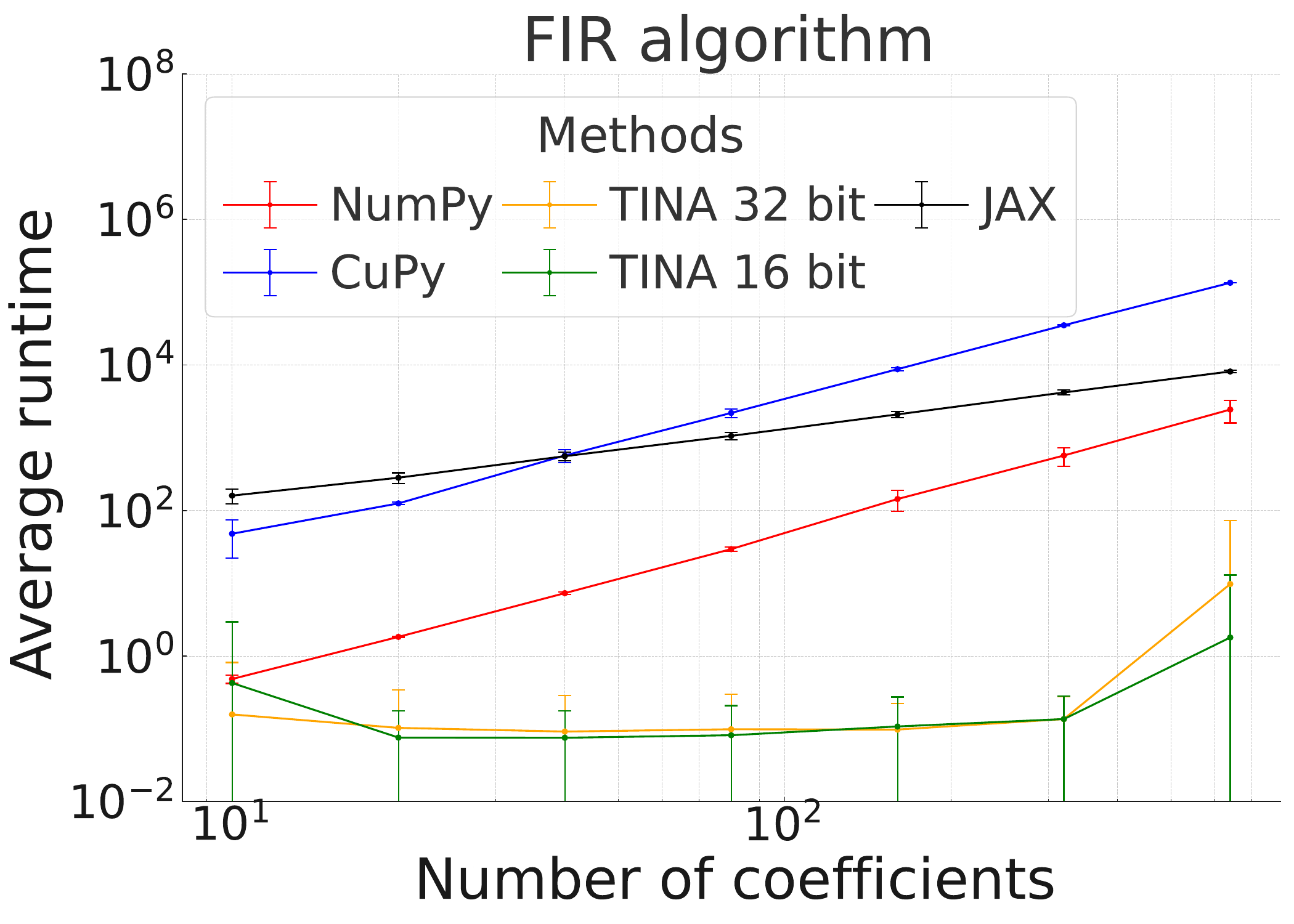}
         \caption{FIR algorithm 
         }
         \label{fig:FIR results}
     \end{subfigure}
     \begin{subfigure}[b]{0.24\textwidth}
         \centering
         \includegraphics[width=1\textwidth]{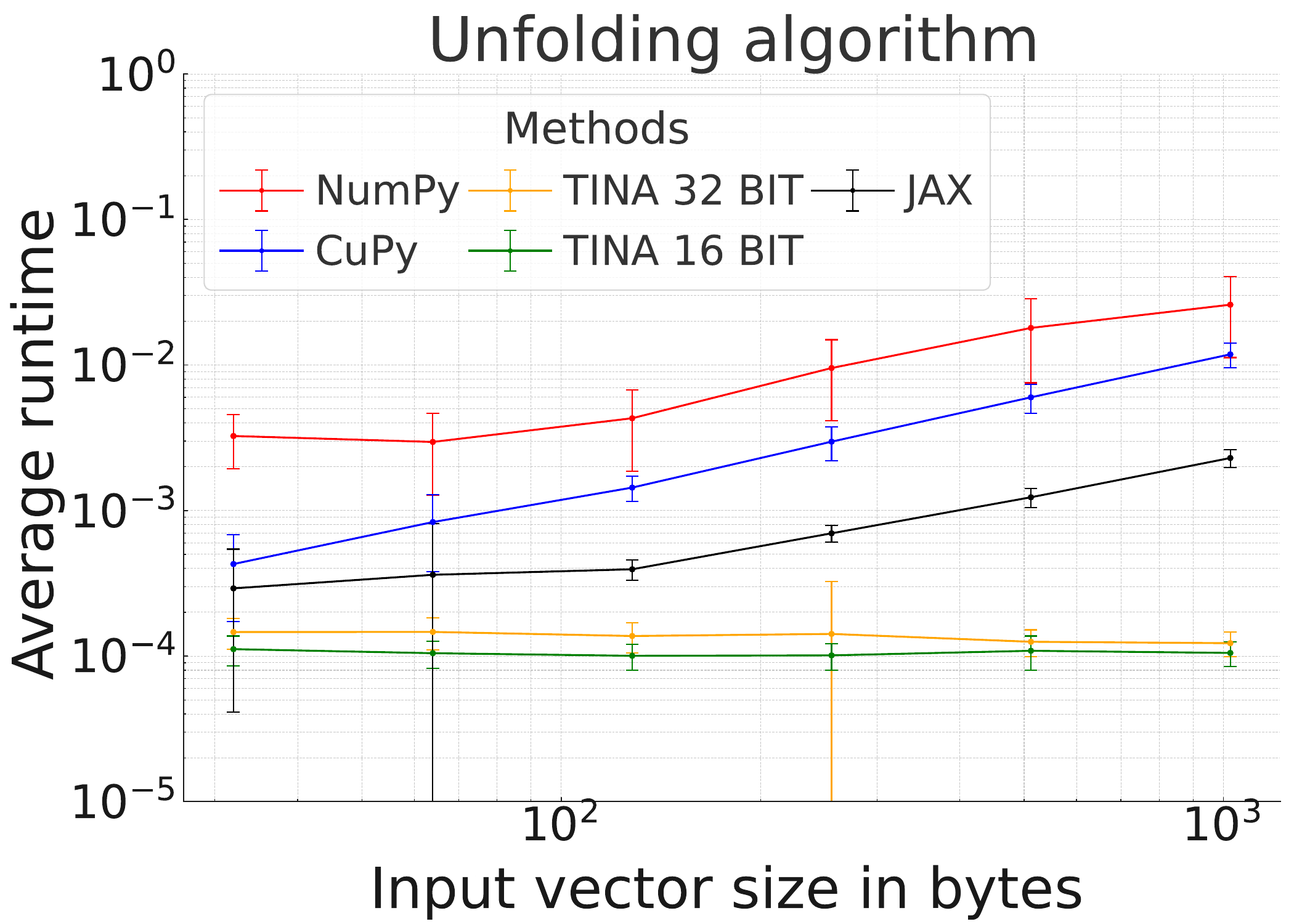}
         \caption{Unfolding algorithm}
         \label{fig:Unfoldingresults}
     \end{subfigure}
        \caption{Runtime of signal processing functions vs input size using TINA, NumPy, CuPy \& JAX}
        \label{fig:signalresults}
\end{figure*}

\begin{figure}[htb]
\vspace*{-0.5cm}
\begin{minipage}[b]{1.0\linewidth}
  \centering
\centerline{\includegraphics[trim={2.8cm 0 0 0},clip, width=9cm]{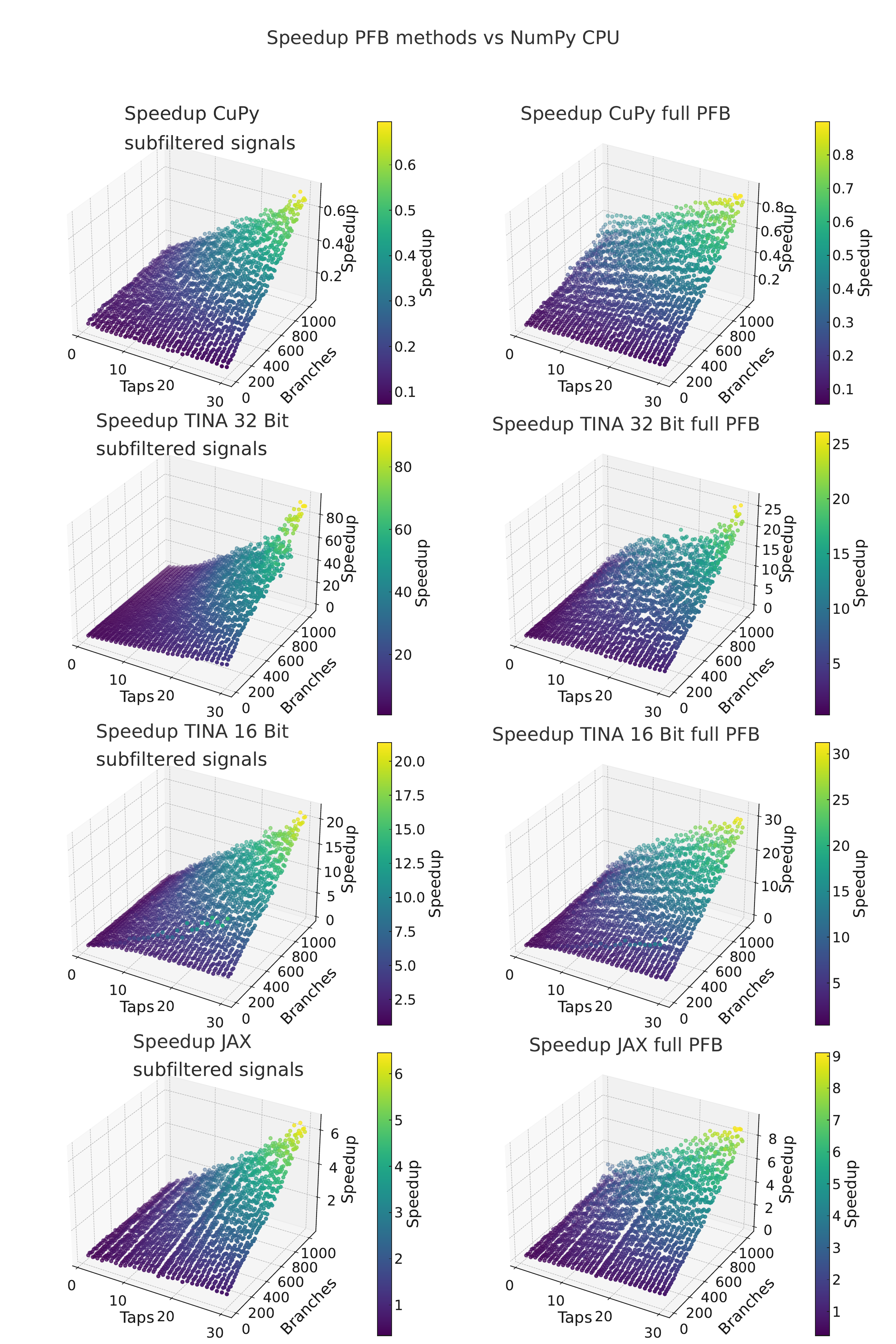}}
\end{minipage}
\caption{GPU speedups with respect to NumPy (CPU) of PFB without (left column) and with (right column) Fourier transform in TINA, JAX \& CuPy}
\label{fig:PFBplots}
\end{figure}

In this section, we discuss our experimental results. \cref{ssec:perf_basic} compares the performance of the basic functions implemented in \cref{sec:OPmapMath} and \cref{sec:OPmapsig} using TINA (using PyTorch running on GPU) vs other solutions, namely NumPy~\cite{2020NumPy-Array} (runs on CPU), CuPy (automatically runs Python on GPU) and JAX (optimized for NN HW). Two results are shown for TINA, with 32 bits (running on GPU CUDA cores) and 16 bits (running on FPU Tensor cores). Then, \cref{sec:PFB} shows the experimental results of a practical use case for implementing a polyphase filter bank in TINA vs other solutions.

The experiments were executed on Google Colab's Tesla T4 GPU (2560 CUDA cores and 320 Tensor cores), alongside an Intel Xeon CPU with 2 vCPUs and 13GB of RAM. By executing our code 100 times and averaging the results, we can show statistically representative performance evaluation. Aside from the TINA 16 bit float implementation, all implementations use 32 bit floating point input data generated randomly. The measurement start once the input data has been copied to the GPU memory in order to minimise the measurement of overhead. 

\subsection{Performance of TINA building blocks} \label{ssec:perf_basic}

\cref{fig:elementmatresults,fig:matmatresults} show that TINA is highly competitive on multiplication-based functions compared to other frameworks resulting in the fastest implementation over a large range of input matrix sizes. For addition-based functions shown in \cref{fig:elementadditionresults,fig:sumresults},  TINA (while competitive) is not as fast as CuPy. This could be attributed to the simplicity of matrix addition functions making it easier for CuPy to find an optimized implementation.

When comparing TINA implementations of DFT and IDFT 
to implementations in other frameworks as shown in \cref{fig:DFTResults,fig:IDFTResults}, we see that JAX appears to be the fastest, followed by TINA. Here, CuPy shows less competitive performance.  
Finally, looking at more computationally complex functions with loops, such as FIR and unfolding  
as shown in \cref{fig:FIR results,fig:Unfoldingresults},  
we can see that TINA is orders of magnitude faster than other frameworks. This is due to the fact that TINA can map functions with data independent loop iterations much more efficiently than other frameworks. 

One important limitation in the TINA framework indicated by our experiment is that the main limiting factor of running operations with larger input vectors or matrices is the large amount of GPU memory used by the TINA framework as we ran into errors once we increased our input sizes beyond what is shown in \cref{fig:signalresults,fig:arithresults}.

\subsection{Use case: polyphase filter bank} 
\label{sec:PFB}

A polyphase filter bank (PFB) channelizes time domain digitized input signals to frequency channels. It is a powerful filter operation that allows for the efficient division of signals into multiple frequency bands, enabling filtering of specific frequency components and simultaneous processing of different frequency components. The PFB is found in many different application domains such as radio astronomy \cite{LOFAR, Price2021ChapterAstronomy}, wireless communication \cite{Harris-telecom}, radar \cite{radar-pfb}, ultrasound imaging \cite{ultrasound-pfb} and quantum computing \cite{PfauFiguliBaehr2018_1000081120}. A PFB receives as input a signal $x(n)$ decomposed into $P$ branches, denoted as $x_p(n')$ which it turns into a set of subfiltered signals $y_p(n')$ \newline
\begin{equation}
y_{p}(n')=\sum_{m=0}^{M-1}h_{p}(m)x_{p}(n'-m)
\label{math:math1}
\end{equation}

where $h_p$ are filter coefficients (otherwise known as taps) that have been divided between the $P$ branches, and $M$ is the number of taps. The subfiltered signals $y_p(n')$ then will be run through either a Discrete or Fast Fourier Transform resulting in the output signals. We can implement this operation by combining multiple FIR filters as detailed in \cref{sec:FIRfilt} and have the resulting output go through a DFT filter as described in \cref{sec:DFT}.

\cref{fig:PFBplots} shows the speedup results of accelerating the subfiltered signals (left column) and PFB (right column) using CuPy, TINA 32 bit, TINA 16 bit, as well as JAX (on GPU) as compared to a naive implementation written in NumPy (on CPU). The figure shows that TINA 32 bit achieves significant acceleration of 25x to 80x \cite{Price2021ChapterAstronomy,Price_2020}, followed by TINA 16 bit with 20x to 30x speedup, while JAX is the next fastest alternative with 6x to 8x speedup. This shows the significant potential of TINA as a portable accelerator language for signal processing applications.

\section{Conclusion}
\label{sec:conclusion}

This paper presented TINA, a novel framework enabling the development of signal processing applications exclusively through the utilization of convolutions and fully connected layers. The paper showed how to map various functions to NN layers and also demonstrated the potential of this approach. Results show that TINA is able to achieve up to 80x GPU speedup over a naive CPU baseline with NumPy \cite{Price2021ChapterAstronomy,Price_2020} for a PFB algorithm compared to 8x speedup achieved by alternative frameworks.

The main drawback currently found by running TINA layers on a GPU is the large amount of memory needed to run the algorithms. Even though for this paper TINA was run on a server-grade GPU, we found the amount of GPU memory to be a limiting factor in TINA's capacity to process larger applications.

Even though these results are promising, the current possible application domains are rather narrow. In order to make TINA more generalisable, further research is needed into mapping more non-NN operations into TINA layers, in addition to investigating ways to make these TINA layers perform efficiently. Furthermore, 
we will investigate adding more TINA layers (linear, recurrent, etc.) to increase the flexibility of the framework, while keeping in mind the three TINA objectives, namely portability, accessibility, and performance.

\subsubsection*{Acknowledgements}
This project has received funding from the Eureka Xecs TASTI project (grant no.~2022005), Horizon Europe research and innovation program RADIOBLOCKS project (grant no.~101093934) and the Netherlands eScience Center RECRUIT project. We would also like to thank Christos Strydis, Lennart Landsmeer and Anna Mészáros for help with reviewing the paper and the mathematics.

{\footnotesize
\bibliographystyle{IEEEbib}
\bibliography{strings,refs}
}

\end{document}